\documentclass[12pt,preprint,aps,pra,groupedaddress,superscriptaddress,amsmath,amssymb,showpacs,showkeys,english]{revtex4-2}
\usepackage{graphicx}
\usepackage{dcolumn}
\usepackage{bm}
\usepackage[T1]{fontenc}
\usepackage{physics}
\usepackage{textcomp}
\usepackage{mathrsfs}
\usepackage{amsmath}
\usepackage{bbm}
\usepackage[normalem]{ulem}
\usepackage{amsbsy}
\usepackage{braket}
\usepackage{amstext}
\usepackage{amssymb}
\usepackage{setspace}
\usepackage{esint}
\usepackage{xcolor,colortbl}
\usepackage[colorlinks,citecolor=blue,urlcolor=blue,linkcolor=blue,hypertexnames=true]{hyperref}
\usepackage{babel}
\usepackage{booktabs}
\usepackage{tikz}
\usepackage{mciteplus}
\begin{document}

\title{Can the present be the average of the future?}

\author{Z. Gedik}

\email{gedik@sabanciuniv.edu}

\affiliation{Faculty of Engineering and Natural Sciences, Sabanci University, Tuzla, Istanbul 34956, Turkey}

\date{\today}

\begin{abstract}
We introduce a two-state vector formalism of quantum mechanics by generalizing Bell’s hidden variable model to higher dimensions and by attributing a physical significance (a state evolving backward in time) to the hidden variable. A simple deterministic and time-symmetric rule for measurement outcomes allows us to obtain the Born rule. It turns out that probabilistic outcomes can be derived from a deterministic assignment and averaging over future states that propagate backward to the present. The assignment rule provides an alternative statement and demonstration of the Pusey – Barrett – Rudolph theorem. 
\end{abstract}


\vspace*{-1.5cm}

\maketitle

\newpage

\section{Introduction}

One of the most fundamental differences between classical and quantum mechanics is the probabilistic nature of the latter. The statistical formalism of quantum theory differs radically from classical statistical mechanics. In the classical case, a fully deterministic theory of the investigated system is at least in principle possible by taking into account all fine-scale parameters. In analogy to the ignored parameters in classical statistical mechanics, the hidden variables were introduced to reproduce stochastic behavior of quantum systems. A proof by von Neumann was thought to rule out hidden variables \cite{vonNeumann}. Hermann \cite{Hermann} figured out the simple error in the proof but it was Bell \cite{Bell} who not only pinpointed the error but also constructed a hidden variable model for a two-level system explicitly. The Born rule provides a link between the mathematical formalism of quantum theory and experiment \cite{Born}. In this work, by reformulating and generalizing the Bell's hidden variable model, we show that it is possible to obtain Born rule from a deterministic formulation of quantum measurement. In other words, it is possible to embed quantum probabilities into a time-symmetric ontology.

\section{Bell's Hidden Variable Model}

After figuring out the error in the proof of von Neumann, Bell constructed probably the simplest hidden variable model \cite{Bell}. Mermin further simplified the construction \cite{Mermin}. The basic idea of the model is, a two-level system, for example a spin-1/2 particle, is described by two (\textbf{m} and \textbf{n}), rather than one (\textbf{m} only), state vectors. Here, \textbf{m} and \textbf{n} denote the Bloch vectors of the pure states \(\ket{\mathbf{m}}\) and \(\ket{\mathbf{n}}\), respectively. According to the model, if we ask, after a measurement, whether this system will be found in state \(\ket{\mathbf{a}}\) (with Bloch vector \textbf{a}), the answer is affirmative if 
\(\textbf{m}\cdot\textbf{a}+\textbf{n}\cdot\textbf{a}> 0\) and otherwise negative (if the expression is less than zero). Here, for simplicity we neglect the equality case since the weight of vanishing left hand side can be statistically neglected. The condition can be rewritten in terms of the state vectors and inner products as 
\begin{equation}
\abs{\braket{\mathbf{m}|\mathbf{a}}}^2 + \abs{\braket{\mathbf{n}|\mathbf{a}}}^2 > 1.
\label{Bell}
\end{equation}
The main idea of Bell's model is to reproduce the predictions of quantum mechanics, namely obtaining \(\abs{\braket{\mathbf{m}|\mathbf{a}}}^2\) as the probability of finding the system in state \(\ket{\mathbf{a}}\), after averaging over \textbf{n}. This average is evaluated simply by integrating over solid angle determined by \textbf{n} on the Bloch sphere. In other words the differential is \(2\pi d(\cos\theta)\), where \(\theta\) is the polar angle. We note that, since \(\abs{\braket{\mathbf{n}|\mathbf{a}}}^2=\cos^2(\theta/2)\), in averaging process we can equivalently use the differential \(d(\abs{\braket{\mathbf{n}|\mathbf{a}}}^2)\).
As far as spinors or, in general, two-state systems are concerned, there is no difference between the two representations. However, the latter allows us to generalize the rule to higher dimensions. This observation and the condition given by Equation (\ref{Bell}) allow us to obtain Born rule in a very simple way. Let \(p=\abs{\braket{\mathbf{m}|\mathbf{a}}}^2
\) and \(q=\abs{\braket{\mathbf{n}|\mathbf{a}}}^2
\). Clearly, \(p\) and \(q\) are real numbers in the interval \([0,1]\). In analogy to Bell's model, assuming that all \(q\) values have equal weight, for a given \(p\) value, the ratio of the \((p,q)\) pairs satisfying Equation (\ref{Bell}) to all possible \((p,q)\) pairs becomes \(p\) which is nothing but the Born rule.

It is clear that in this derivation uniform \(q\) distribution plays a crucial role. For example, we can write the backward evolving state in terms of generalized Gell-Mann matrices and introduce a uniform distribution on the generalized Bloch sphere. However, in this case, starting from three dimensions, it is known that not all points on the Bloch sphere correspond to a physical state. More precisely, traces of the first, second and third powers of the density matrix must be unity and in general this leads a complicated set of equations \cite{Samuel}. Another possibility is to scan the Hilbert space by using (Haar) unitary random matrices \cite{Mehta}. Rather than examining possible averaging procedures, our sole aim is to show the existence of a simple distribution giving the Born rule. 

\section{Higher Dimensions and Time Symmetry}

Bell did not attribute any physical significance to the hidden or additional parameters. His sole aim was to demonstrate that probabilistic predictions of quantum mechanics can be reproduced in a deterministic way by introducing additional parameters. Equation (\ref{Bell}) can be generalized to higher dimensions by replacing \(\ket{\mathbf{m}}\) and \(\ket{\mathbf{n}}\) with arbitrary states so that the rule becomes

\begin{equation}
\abs{\braket{\Psi_\uparrow|a}}^2 + \abs{\braket{\Psi_\downarrow|a}}^2 > 1.
\label{Gedik}
\end{equation}

Here, in generalization to arbitrary dimension, we replace \(\ket{\mathbf{m}}\) and \(\ket{\mathbf{n}}\) by \(\ket{\Psi_\uparrow}\) and \(\ket{\Psi_\downarrow}\), respectively, and \(\ket{a}\) is the state that we wonder if the system will be  found in after the measurement. Up and down arrows indicate quantum states propagating forward and backward in time. This is one further step, an additional assumption that we include into Bell's model. In this way, the rule becomes fully time-symmetric in a very natural way. 

Equation (\ref{Gedik}) requires a consistency check. We will show that if \(\ket{a}\) and \(\ket{b}\) are two orthogonal states, for example two eigenstates of a Hermitian operator, then rule (\ref{Gedik}) cannot be satisfied by both states simultaneously. Since, \(\ket{a}\) and \(\ket{b}\) are two orthogonal states, we have \(\abs{\braket{\Psi_\uparrow|a}}^2 + \abs{\braket{\Psi_\uparrow|b}}^2\le 1\) and \(\abs{\braket{\Psi_\downarrow|a}}^2 + \abs{\braket{\Psi_\downarrow|b}}^2\le 1\).  Summing the two inequalities side by side, we obtain the required result: Assuming both \(\ket{a}\) and \(\ket{b}\) satisfy (\ref{Gedik}) gives a sum greater than 2, while orthogonality forces the same sum less than or equal to 2. 

Given a pair of states \(\ket{\Psi_\uparrow}\) and \(\ket{\Psi_\downarrow}\), it might turn out that none of the basis vectors \(\ket{a}\) satisfies the rule (\ref{Gedik}) but if there is a solution to the inequality, it is unique. There might be other pairs, say \(\ket{\Phi_\uparrow}\) and \(\ket{\Phi_\downarrow}\), satisfying the rule (\ref{Gedik}), but \((\ket{\Psi_\uparrow}\),\(\ket{\Psi_\downarrow}\)) and \((\ket{\Psi_\downarrow}\),\(\ket{\Psi_\uparrow}\)) will always give the same results which is a manifestation of the time symmetry. Time reversal operation is nothing but the exchange of forward and backward evolving states. 

It is possible to generalize (\ref{Gedik}) to include mixed states as
\begin{equation}
\operatorname{Tr}[(\rho_\uparrow+\rho_\downarrow)\Pi_a]> 1
\label{GGedik}
\end{equation}
where \(\Pi_a=\ketbra{a}{a}\) is the projector corresponding to the measurement while \(\rho_\uparrow\) and \(\rho_\downarrow\) are the density matrices describing states evolving forward and backward in time, respectively. Time evolution is governed by \(U^\dagger\rho_\uparrow U +U\rho_\downarrow U^\dagger\) where \(U\) is the usual time evolution operator. In analogy to the time-independent Schrödinger equation, we can obtain an expression satisfied by stationary states. For a time independent Hamiltonian \(H\), invariance of \(\rho_\uparrow+\rho_\downarrow\) after an infinitesimal time evolution gives the condition 
\begin{equation}
[\rho_\uparrow,H]=[\rho_\downarrow,H]. 
\label{stat}
\end{equation}
Therefore, stationary state problem is reduced to finding density matrices \(\rho\) having the same commutator with a given hermitian matrix, namely the Hamiltonian. Let this commutator be \(K\). It is easy to see that \(K\) must have zero diagonal in the eigenbasis of \(H\) and for non-diagonal elements \(i\ne j\) the equation \([\rho,H]=K\) gives \(\rho_{ij}=K_{ij}/(E_i-E_j)\) where \(E_i\) and \(E_j\) are the corresponding eigenvalues of the Hamiltonian. Diagonal elements \(\rho_{ii}\) are completely free (other than being real, adding up to unity and leading to non-negative eigenvalues). We note that inside degenerate subspaces, \(K\) must vanish. Equations (\ref{GGedik}) and (\ref{stat}) show that while the sum of forward and backward evolving density matrices determine the measurement outcome, their difference plays role in time-dependence of the system. Stationary states are the ones where the difference commutes with the Hamiltonian. 

The idea of time-symmetric formulation of quantum mechanics, or more precisely of quantum measurement is quite an old problem \cite{Watanabe,ABL}. We note that, the two state formalism introduced here is different from the one in the context of weak values \cite{AAV}. In our case, there is no post-selection process. The backward-propagating state allows us to predict the result of measurement in a deterministic way. The main idea of weak values is that final state after the measurement is not necessarily an eigenstate of the measured dynamical variable and furthermore, through post-selection, we can obtain different ensembles not available in usual measurements. In our formalism, in the case of strong measurement, both \(\ket{\Psi_\uparrow}\) and \(\ket{\Psi_\downarrow}\) collapse on \(\ket{a}\) and corresponding eigenvalue is \(a\). In the case of weak measurement on the other hand, final state \(\ket{\Phi}\) is arbitrary. Our averaging rule gives the usual probability \(\abs{\braket{\Psi_\uparrow|\Phi}}^2\) for this event and the corresponding weak value is \(\braket{\Phi|A|\Psi_\uparrow}/\braket{\Phi|\Psi_\uparrow}\) \cite{AAV}.

\section{Elements of Physical Reality and Pusey – Barrett – Rudolph Theorem}

According to Einstein, Podolsky and Rosen (EPR), if we can predict with certainty the value of a physical quantity, then there exists an element of physical reality corresponding to this physical quantity \cite{EPR}. Rule (\ref{Gedik}) makes predictions about the measurement outcomes with probability equal to unity and hence they correspond to elements of physical reality in the EPR sense. 

Symmetric informationally complete quantum measurements (SICPOVMs) are a set of projectors that can be used as basis to expand any density matrix \cite{ZaunerT, SICPOVM, Zauner}. SICPOVM operators \(\{\Pi_k\}\) are equiangular in the sense that \(tr(\Pi_k\Pi_l)=1/(d+1)\) for any pair of \(\Pi_k\) and \(\Pi_l\) with \(k\neq l\). Here, \(d\) is the dimensionality of the space. According to Zauner's conjecture, there exist \(d^2\) such projectors in all dimensions \cite{ZaunerT}. 

Using SICPOVMs we can write the density operators in (\ref{GGedik}) as \(\rho_\uparrow=\sum_k \lambda_{\uparrow k}\Pi_k\) and \(\rho_\downarrow=\sum_k \lambda_{\downarrow k}\Pi_k\). We note that, since \(\operatorname{Tr}\rho_\uparrow=\operatorname{Tr}\rho_\downarrow=1\), the expansion coefficients satisfy \(\sum_k \lambda_{\downarrow k}=\sum_k \lambda_{\uparrow k}=1\). More precisely, we can show that \(-1/d\le\lambda_{\downarrow k},\lambda_{\uparrow k}\le1\). For simplicity, we denote the sum of the density matrices as \(\rho=\rho_\uparrow+\rho_\downarrow=\sum_k \lambda_k\Pi_k\), therefore \(\sum_k \lambda_k=2\). Now, as a special case of rule (\ref{GGedik}), let us ask what if \(\Pi_a\) is one of the SICPOVM operators? A simple calculation shows that the expansion coefficient should satisfy the condition \(\lambda_k> 1-1/d\). In this case, we can find more than one solutions. For example in \(d=2\), the set \{3/4, 3/4, 1/4, 1/4\} contains two coefficients satisfying the condition \(\lambda_k> 1/2\). This does not contradict our earlier result for orthogonal states since SICPOVM states are not mutually exclusive. 

In their analysis on the reality of the wave function Pusey, Barrett, and Rudolph (PBR) demonstrated that non-orthogonal states have different underlying physical realities \cite{PBR}. In our formalism, we can ask a related question as follows: Given two states \((\ket{\Psi_\uparrow}\),\(\ket{\Psi_\downarrow}\)) and \((\ket{\Phi_\uparrow}\),\(\ket{\Phi_\downarrow}\)), is there a measurement, namely a state \(\mid a\rangle\), for which inequality (\ref{Gedik}) is satisfied by one of them, while it is violated by the other? For simplicity, let us first consider the spin-1/2 case with \(\ket{\Psi_\uparrow}=\ket{\mathbf{z}}\),\(\ket{\Psi_\downarrow}=\ket{\mathbf{z}^{\prime} }\), \(\ket{\Phi_\uparrow}=\ket{\mathbf{x}}\), and \(\ket{\Phi_\downarrow}=\ket{\mathbf{x}^{\prime} }\) where \(\ket{\mathbf{z}}\) (\(\ket{0}\) in PBR) and \(\ket{\mathbf{x}}\) (\(\ket{+}\) in PBR) denote up-spin states along \(z-\) and \(x-\) axes, respectively. In the plane determined by the vectors \(\mathbf{z}+\mathbf{z}^{\prime}\) and \(\mathbf{x}+\mathbf{x}^{\prime}\), we can always find a vector \textbf{a} which has positive projection on \(\mathbf{z}+\mathbf{z}^{\prime}\), and a negative projection on \(\mathbf{x}+\mathbf{x}^{\prime}\). Therefore, it is possible to distinguish \((\ket{\mathbf{z}},\ket{\mathbf{z}^{\prime} })\) and \((\ket{\mathbf{x}},\ket{\mathbf{x}^{\prime} })\) with a single measurement.

The SICPOVM formalism mentioned in the discussion of EPR elements of reality, can also be used in more general form of the PBR argument. Now, instead of \((\ket{\mathbf{z}},\ket{\mathbf{z}^{\prime} })\) and \((\ket{\mathbf{x}},\ket{\mathbf{x}^{\prime} })\), let us consider any pair of quantum states 
\((\ket{\Psi_{0\uparrow}},\ket{\Psi_{0\downarrow}})\) (\(\ket{\Psi_0}\) in PBR) and 
\((\ket{\Psi_{1\uparrow}},\ket{\Psi_{1\downarrow}})\) (\(\ket{\Psi_1}\) in PBR). Since the two states are assumed to be different, \(\lambda_k\) coefficients satisfying (\ref{GGedik}) will also be different, say \(\lambda_{k_0}\) and \(\lambda_{k_1}\), respectively. This means that,  a single \(\Pi_{k_0}\) or \(\Pi_{k_1}\) measurement can distinguish between \(\ket{\Psi_0}\) and \(\ket{\Psi_1}\). Hence, our hidden-variable model provides a demonstration of the PBR theorem for a special set of quantum states.   

\section{Conclusion}

In analogy to Bell's hidden variable model for two-level systems, we introduced a  two-state vector formalism which can be applied to quantum systems in arbitrary dimensions. A fully time-symmetric, simple rule can be used to obtain Born rule. In this time-symmetric hidden-state framework, Born probabilities emerge from our ignorance about the backward states.  Uncertainty in the description of past events which is analogous to the uncertainty in the prediction of future events is in the very nature of quantum mechanics \cite{ETP}. The entropic direction of time, imposed by the macroscopic measurement system, requires an averaging over states evolving backward in time.

\section{Acknowledgements}

Part of this work was carried out at Bilimler Köyü in Foça. The author gratefully acknowledges useful and stimulating discussions with  A. Dizdar, A. Kabakçıoğlu, and G. Karpat.


\begin{thebibliography}{4}

\bibitem{vonNeumann}
J. von Neumann, Mathematische Grundlagen der Quantenmechanik (Springer, Berlin, 1932).  

\bibitem{Hermann} G. Hermann, Determinismus und Quantenmechanik (1933); reprinted in K. Herrmann, ed., Grete Henry-Hermann: Philosophie–Mathematik–Quantenmechanik, Springer (2019), p. 185.

\bibitem{Bell}
J. S. Bell, On the problem of hidden variables in quantum theory, Reviews of Modern Physics \textbf{38}, 447 - 452 (1966).

\bibitem{Born}
M. Born, Zur Quantenmechanik der Stoßvorgänge, Zeitschrift für Physik \textbf{37}, 863 – 867, (1926).  

\bibitem{Mermin}
N. D. Mermin, Hidden variables and two theorems of John Bell, Reviews of Modern Physics \textbf{65}, 803 - 815 (1993).

\bibitem{Samuel} S. B. Samuel and Z. Gedik, Group theoretical classification of
SIC-POVMs, Journal of Physics A: Mathematical and Theoretical \textbf{57}, 295304 (2024).

\bibitem{Mehta} M. L. Mehta, Random Matrices, Academic Press (1991).

\bibitem{Watanabe}
S. Watanabe, Symmetry of Physical Laws. Part III. Prediction and Retrodiction, Reviews of Modern Physics \textbf{27}, 179 - 186 (1955).

\bibitem{ABL} Y. Aharonov, P. G. Bergmann, and J. L. Lebowitz, Time Symmetry in the Quantum Process of Measurement, Physical Review \textbf{134} B1410 - B1416 (1964).

\bibitem{AAV} Y. Aharonov, D. Z. Albert, and L. Vaidman, How the result of a measurement of a component of the spin of a spin-1/2 particle can turn out to be 100, Physical Review Letters \textbf{60}, 1351 – 1354 (1988). 

\bibitem{PBR} M. F. Pusey, J. Barrett, and T. Rudolph, On the reality of the quantum state, Nature Physics \textbf{8}, 475 – 478 (2012). 

\bibitem{EPR} A. Einstein, B. Podolsky, and N. Rosen, Can quantum-mechanical description of physical reality be considered complete?, Physical Review \textbf{47}, 777 – 780 (1935). 

\bibitem{ZaunerT} G. Zauner, Quantendesigns: Grundzüge einer nichtkommutativen Designtheorie, Ph.D. Thesis, University of Vienna, (1999). 

\bibitem{SICPOVM} J. M. Renes, R. Blume-Kohout, A. J. Scott, and C. M. Caves, Symmetric informationally complete quantum measurements, Journal of Mathematical Physics \textbf{45}, 2171 – 2180 (2004). 

\bibitem{Zauner} G. Zauner, Quantum designs: Foundations of a noncommutative design theory, International Journal of Quantum Information \textbf{9}, 445 – 507 (2011).

\bibitem{ETP} A. Einstein, R. C. Tolman, and B. Podolsky, Knowledge of past and future in quantum mechanics, Physical Review \textbf{37}, 780 – 781 (1931).

\end{thebibliography}
\end{document}